\documentclass[
  journal=largetwo,
  manuscript=psychology,
  year=2023,
  volume=37,
]{cup-journal}

\usepackage{amsmath}
\usepackage[nopatch]{microtype}
\usepackage{booktabs}
\usepackage{pgfplots}
\pgfplotsset{width=7cm,compat=newest}

\title{Social AI Improves Well-Being Among Female Young Adults}

\author{Xiaoding Lu}
\affiliation{Special thanks to Isis Buse for her contribution in writing the article}

\author{Ebony Zhang}
\affiliation{University College London}

\usetikzlibrary{patterns}
\begin{document}

\begin{abstract}
The recent growth of AI language models has brought Social AI to the fore as a new means of entertainment and interaction, particularly for teenagers and young adults. This paper investigates the impact of interactions with generative AI agents on users’ mental and social health and well-being; a subject which has generated much debate both within academic communities and throughout the public. Our study, based on a survey of 5,260 users of Chai AI, a Social AI platform, indicates the considerable benefits of Social AI on user mental health, while also revealing a notable variance in responses across demographic strata. Female users reported the most substantial benefits from engagement with Social AI: 43.4\% strongly agreed that Social AI positively impacted their mental health, while 10.5\% fewer of their male counterparts agreed with this statement. A further 38.9\% of female users strongly agreed that Social AI made their anxiety more manageable, compared with 30.0\% of male users and 27.1\% of other genders. New media and technological advancements have frequently been the subject of moral and ethical scrutiny, and the advent of Social AI has been no different. The findings of research conducted within this paper demonstrate the importance of an evidence-based approach in discussions surrounding the behavioural and social impacts of these emerging technologies. \end{abstract}

\noindent The impacts of traditional forms of social media on user mental health and well-being have been the subject of extensive research and discussion. Recent developments in the field of AI language models have accelerated the growth of Social AI, platforms which allow users to connect and interact with AI-generated characters, and have brought these new platforms into existing debates concerning social and mental well-being.

Widespread debate surrounding the potentially negative impacts of excessive screen time and social media usage has given way to a abundance of academic research and literature. One such study, based on a comprehensive meta-analysis, encompassing 37 effect sizes from 33 separate studies published between 2015 and 2019, concluded that the usage and extent of screen time had minimal impacts on mental health, including suicidal thoughts \autocite{Ferguson2022}. These findings are demonstrative of a general trend surrounding discussions of this subject, in which initial assumptions of an overwhelmingly negative impact of social media on mental health are often reassessed and contested through evidence-based analyses.

Social AI, while occupying a similar role to traditional forms of social media by enabling forms of social interaction, possess unique elements. While traditional forms of social media facilitate interaction between human individuals or groups, Social AI operates through interactions between humans and artificially-generated characters. This new and distinct dimension of interaction introduced by AI has been scrutinised over possible negative effects on users' mental health, social anxiety, and behavioural development. This paper aims to explore whether the concerns associated with the use of traditional social media platforms are relevant to analyses of the impact of Social AI on mental and social health across social and demographic strata. 

While there is an existing body of well-developed research and academic literature analysing the impacts of traditional social media on user mental health, there has been minimal research analysing the effect of Social AI on health and well-being. This paper hopes to draw upon, expand, and supplement this existing bank of research in order to understand the extent to which the concerns and risks generally associated with use of social media can, and should, be applied to this new social medium, Social AI. In dis-aggregating data across different demographic strata, including gender and age, we hope to provide an in-depth analysis of the impacts of Social AI on user mental health and social well-being. 

This paper is based upon evidence and data analysed from the survey responses of 5,260 users of the Chai AI platform. By providing direct indicators of user experience and impact, evidence collected from users is essential in managing and reassessing general assumptions regarding the impacts that these platforms have on mental and social health. As Social AI provides a new platform for social interactions, an analysis of user experience will both further and expand ongoing discussions concerning the roles of technology in shaping mental health, and social and behavioural attitudes.

\section{Traditional Social Media and Its Effect on Mental Health}

The exponential growth of social media as a leading method of social interaction over the past decade has led to widespread debate over its impacts on mental health. Scholars, policymakers, and members of the public have dedicated considerable energy to analyses concerning the ability of technology to help, or hinder, positive mental health and behavioural development. The concept of 'moral panic', a term for the significant anxiety which has developed alongside the rise of these platforms and their potential for negative impacts, has been at the heart of this discourse \autocite{Orben2020}. Certain scholars and studies, however, have disputed the existence and validity of the data or evidence that these anxieties are based upon \autocite{Bowman2015, Cantor2009, Ferguson2013, ODonohue1996}. Others have asserted that this tendency towards 'moral panic' often diminishes as growing evidence to the contrary eases concerns over potentially negative outcomes \autocite{Bowman2015}; \autocite{Markey2015}. Debates concerning fears over the relationship between increased screen time and poor mental health, particularly among young people, provide an apt example of this pattern.

This tendency towards a social 'moral panic' at the evolution of new forms of entertainment consumption has been consistently experienced throughout modern history \autocite{Orben2020}. These concerns which once surrounded the use of radio \autocite{Preston1941}, were then projected onto comic books \autocite{Wertham1954}, and dominated discussions about extensive consumption of television content in the 1980s \autocite{Rubinstein1983}. These examples show the general trend of immediate panic, associated with perceived risks of new and unknown platforms, which are then tempered over time, and eventually conferred to the newest form of entertainment or media. As a novel platform for social interaction, the  panic currently surrounding the use of Social AI appears to be overwhelmingly similar.

Current debates regarding the impact of screen time on mental health are strongly related to concerns over both social media and Social AI usage. One notable aspect of discussion has been widespread concern over the impact of excessive screen time on young women and girls, in particular \autocite{CDC2020WISQARS, Twenge2017Smartphones, Orben2019, Twenge2020}. However, existing data on the subject indicates that the highest increase in suicide rates in the United States over the past two decades are documented to be among middle-aged men \autocite{CDC2020WISQARS}. Equally, global data does not consistently or homogeneously support a correlation between increased screen time and increased suicide rates, with some countries even exhibiting a decrease in suicide rates over recent years \autocite{Eurostat2020SuicideRate}. These conclusions are indicative that these claims and concerned should be treated with caution, and tested through an evidence-based analysis, which may offer alternative or supplementary conclusions to the initial assumption. 

Studies and research analyses have attempted to both correlate and absolve the extent of screen time with poor mental health outcomes, and often offer conflicting conclusions \autocite{DennisonFarris2017, Ferguson2017, Huang2017, Kleppang2017, Chan2013, Park2016}. The principal reason for these inconsistencies are the challenges associated with differing definitions and evolving understandings of what constitutes 'screen time'. Additionally, methodology poses further difficulties, due to issues concerning potential for coincidental relationships, misunderstood causality, and misuse of unstandardised methods \autocite{Drummond2020, McDonnell2019, Want2014, Whyte2016, Elson2014, Lykken1968}. 

Practical difficulties in analysing the impact of different forms of social media, including Social AI, coupled with the emerging 'moral panic' punctuating the current discourse highlights the need to approach this topic with an analysis thoroughly based on evidence and data.

\section{AI Language Models and Social AI}

The recent development of AI language models have generated significant growth in the investment and interest in, and use of artificial intelligence. ChatGPT, which generates written content based on prompts, questions, and directions through user interaction, has transformed the rate and form of human interaction with AI. The sophisticated, and increasingly human-like text produced by ChatGPT is founded upon various large-scale machine learning algorithms, particularly transformer-based models \autocite{NIPS2017_3f5ee243, NEURIPS2020_1457c0d6}.

These advancements in the field of language models have propelled the use and popularity of artificial companionship. These AI companions are designed to mimic human interaction, and are characterised by astute response-memory, empathy, and even specific character traits ascribed to different bots, representing a transition from the traditionally functional and task-oriented use of AI. AI companionship is one aspect of the broader umbrella term of Social AI. Some platforms, such as Replika and Pi, operate specifically to fulfil the role of companion, where interaction is directed toward a singular AI entity, promoting companionship specifically. Others, such as Chai AI and Character AI fit into more diverse categories of social AI, where the chosen AI bot is tailored specifically to needs and directions of different conversations, and can assume the role of companion, but also of teacher, assistant, or debate-partner \autocite{Irvine2022}. In this case, companionship AI becomes one facet of Social AI. 

In line with more general trends surrounding the uptake and engagement with AI, usage of Social AI and AI companionship models has been most prevalent among adults and teenagers across the United States, with a reported average active usage of over one hour per day \autocite{Ghosh2023Replika}. Much like the extensive concerns which have appeared within the discourse surrounding forms of traditional social media, there has been some apprehension as to the impact of Social AI on user mental health.

\begin{figure}[h]
\centering
\begin{tikzpicture}
    \begin{axis}
        [
            title={Mental Health Issues Across Gender},
            title style={at={(axis description cs:-0.015, 1.035)}, anchor=north west},
            ybar,
            bar width=5pt,
            legend style={at={(0.5, 0.97)},
            anchor=north,legend columns=-1},
            ylabel={Frequency Density \%},
            xlabel={Likert Scale},
            symbolic x coords={1,2,3,4,5},
            xtick=data,
            xtick pos=left,
            width=0.98\textwidth,
            ymin=0,
            ymax=55,
        ]
        \addplot[pattern=north west lines, pattern color=black, area legend] coordinates {(1, 10.9) (2, 6.92) (3, 25.0) (4, 20.6) (5, 36.5)};
        \addplot[fill=gray, draw=black, area legend] coordinates {(1, 14.8) (2, 8.92) (3, 27.9) (4, 18.0) (5, 30.3)};
        \addplot[fill=white, area legend] coordinates {(1, 6.41) (2, 3.49) (3, 21.0) (4, 19.5) (5, 49.6)};
        \legend{Female,Male,Other}
    \end{axis}
\end{tikzpicture}
\caption{Answer to "I have experienced in the past with mental health issues" Likert scale density breakdown by gender. With 1 being "strongly disagree" and 5 being "strongly agree". N=5,260}
\label{fig:mental_health_experience}
\end{figure}
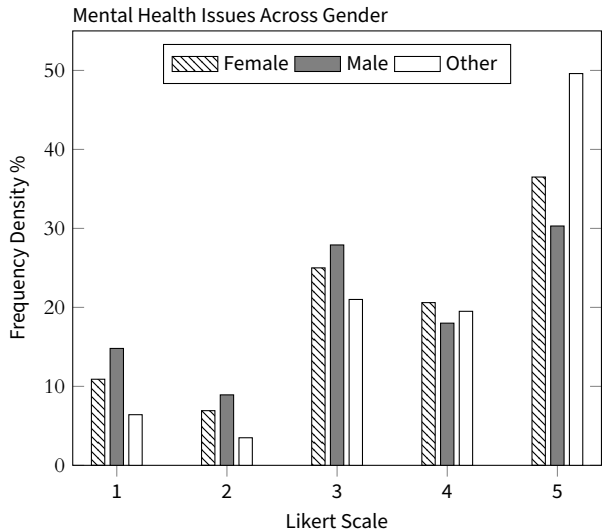

\begin{figure}[h]
\centering
\begin{tikzpicture}
    \begin{axis}
        [
            title={Social Anxiety Across Gender},
            title style={at={(axis description cs:-0.015, 1.035)}, anchor=north west},
            ybar,
            bar width=5pt,
            legend style={at={(0.5, 0.97)},
            anchor=north,legend columns=-1},
            ylabel={Frequency Density \%},
            xlabel={Likert Scale},
            symbolic x coords={1,2,3,4,5},
            xtick=data,
            xtick pos=left,
            width=0.98\textwidth,
            ymin=0,
            ymax=75,
            ytick={10,20,30,40,50,60},
        ]
        \addplot[pattern=north west lines, pattern color=black, area legend] coordinates {(1, 7.1) (2, 3.79) (3, 12.3) (4, 20.6) (5, 56.2)};
        \addplot[fill=gray, draw=black, area legend] coordinates {(1, 10.4) (2, 5.6) (3, 15.9) (4, 22.2) (5, 45.8)};
        \addplot[fill=white, area legend] coordinates {(1, 2.62) (2, 2.04) (3, 7.29) (4, 17.2) (5, 70.8)};
        \legend{Female,Male,Other}
    \end{axis}
\end{tikzpicture}
\caption{Answer to "I have experienced in the past with social anxieties" Likert scale density breakdown by gender. With 1 being "strongly disagree" and 5 being "strongly agree". N=5,260}
\label{fig:social_anxiety_experience}
\end{figure}
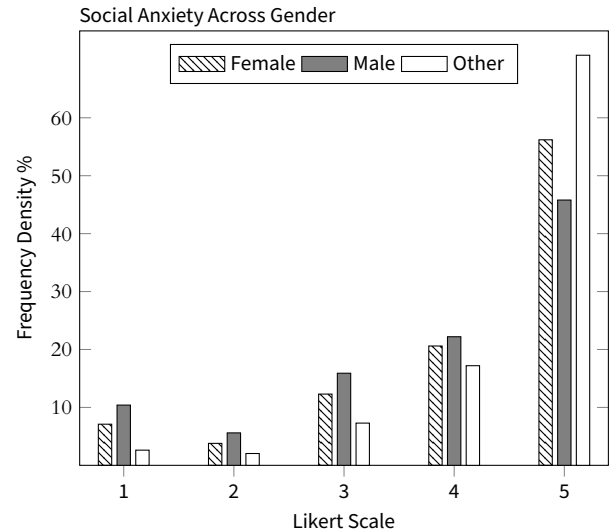

\begin{figure}[h]
    \centering
    \begin{tikzpicture}
        \begin{axis}
            [
                title={Mental Health and Social Anxiety Across Age Groups},
                title style={at={(axis description cs:-0.015, 1.035)}, anchor=north west},
                xlabel={Age Group},
                ylabel={Likert Mean},
                xtick={1,2,3,4},
                xticklabels={$\leq$18,19-24,25-30,$\geq$30},
                ymin=3, ymax=4.5, 
                width=0.98\textwidth,
                legend style={at={(0.5,0.97)},anchor=north,legend columns=-1},
                xtick pos=left,
            ]
            \addplot+[
                black,
                mark=*,
                mark options={fill=black},
                error bars/.cd,
                y dir=both,
                y explicit,
                error bar style={black, line width=1pt},
                error mark options={
                  rotate=90,
                  line width=1pt
                }
            ]
            coordinates {
                (1,3.551097) +- (0,0.031510)
                (2,3.590732) +- (0,0.023971)
                (3,3.715084) +- (0,0.089816)
                (4,3.500000) +- (0,0.120232)
            };
            \addlegendentry{Mental Health Issues}
            
            \addplot+[
                black,
                mark=o,
                mark options={fill=black, solid},
                dashed,
                error bars/.cd,
                y dir=both,
                y explicit,
                error bar style={black, line width=1pt},
                error mark options={
                  solid,
                  rotate=90,
                  line width=1pt
                }
            ]
            coordinates {
                (1,4.061530) +- (0,0.029413)
                (2,4.067077) +- (0,0.022284)
                (3,4.078212) +- (0,0.086710)
                (4,4.023810) +- (0,0.106754)
            };
            \addlegendentry{Social Anxiety}
        \end{axis}
    \end{tikzpicture}
    \caption{Comparison of Likert mean responses to "I have experienced in the past with mental health issues" and "I have experienced social anxiety before" by age group, with 1 standard error shown. A Likert score 3 being baseline, i.e. "No opinion".}
    \label{fig:mental_social_anxiety_age}
\end{figure}
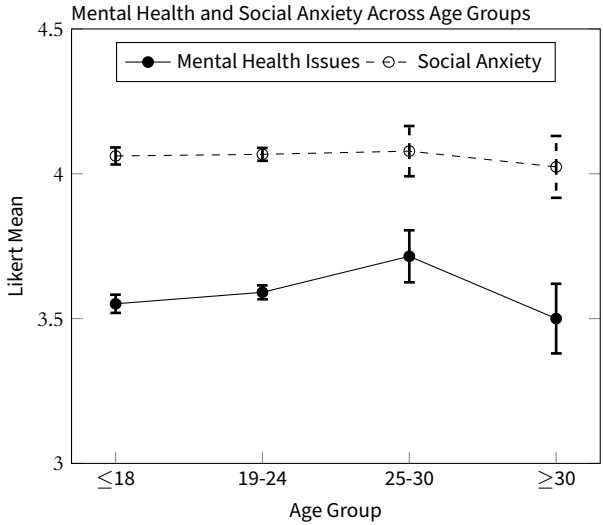

\section{Method and Analysis}
\subsection{Survey Design}

The evidence reflected in this study is based upon a survey of 5,260 individual users of the Chai AI platform. Questions asked by the survey were devised in order to determine the effects of Social AI on user mental health and well being, in order to further analyse the validity of assumed associated risks. 

During the survey, respondents were asked a series of questions designed to give their subjective analysis of their mental health and extent of social anxiety, using the Likert scale \autocite{Jebb2021}, both before and after interacting with the Social AI platform. The survey also asked for basic demographic information, including gender and age. The questions were as follows:

\begin{itemize}
    \item \textbf{Age Group:} ``How old are you?'' with options: $<$18, 19-25, 26-30, $>$30. This was a single-choice question.
    \item \textbf{Gender:} ``What is your gender?'' with options: Male, Female, Other, Prefer not to say. This was also a single-choice question.
    \item \textbf{Past Mental Health Issues:} ``I have experienced in the past with mental health issues.'' Participants rated this statement on a Likert scale from 1 (Strongly Disagree) to 5 (Strongly Agree), with 3 being Neutral.
    \item \textbf{Past Social Anxieties:} ``I have experienced in the past with social anxieties.'' This was similarly rated on a Likert scale from 1 to 5.
    \item \textbf{Impact of Social Chat AI on Mental Health:} ``Social Chat AI has had a positive impact on my mental health.'' Participants rated this statement on the same Likert scale.
    \item \textbf{Impact of Social Chat AI on Social Anxieties:} ``Social Chat AI has had a positive impact on my social anxieties.'' This was also rated on the Likert scale from 1 to 5.
\end{itemize}

\subsection{Data Collection}

The survey was hosted directly on the Chai AI platform over a period of two weeks. By organising data collection from the platform itself, it was ensured that respondents were genuine users of the platform. After collection, the data was anonymized for analysis. 

\subsection{Measuring User Demographics, Mental Health and Social Anxieties}

The survey data, represented in Figures \ref{fig:mental_health_experience}, \ref{fig:social_anxiety_experience}, and \ref{fig:mental_social_anxiety_age}, provides a comprehensive overview of the mental health issues and social anxieties experienced by users across different gender and age groups.

The data collected from survey respondents has been dis-aggregated by both gender identity and age groups in order to more holistically assess the impacts of Social AI on user mental health. The highest proportion of respondents identified as female, at 44.5\%. A further 38\% identified as male, 6.4\% identified as other, and 11.1\% preferred not to say. The proportion of those participants identifying as other genders corresponds well to the national (US) average, which sits at just over 5\% \autocite{Meerwijk2017}. Young adults constituted the majority of survey participants, with 58.4\% being 19-25, followed by 35\% being under 18, 3.8\% 25-30, and 2.9\% over 30 years of age. 

Figure \ref{fig:mental_health_experience} shows a significant degree of variance across genders in reported past mental health issues. The highest rate of strong agreement to having experienced mental health issues was found among users identifying as other genders, at 49.6\%, compared with 36.5\% of those identifying as female and 30.0\% of those identifying as male. Similarly, with a rate of 69.1\%, participants identifying as other genders also reported the strongest level of overall agreement (responses of Strongly Agree and Agree combined) to having experienced mental health issues in the past. Again, this was followed by 57.1\% of those identifying as female, and 48.3\% of those identifying as male. 

Figure \ref{fig:social_anxiety_experience} indicates that those identifying as other genders constituted the highest percentage of strong agreement to having experienced social anxiety (70.8\%), with combined responses of Agree and Strongly Agree at 88.0\%. Those identifying as female exhibited a substantial tendency toward experiencing social anxiety, with their responses of Agree and Strongly Agree combined sitting at 76.8\%. Rates of agreement (Agree and Strongly Agree combined) among those identifying as male were 68.0\%. 

In comparisons across age groups, shown in Figure \ref{fig:mental_social_anxiety_age}, users aged 25-30 report a slightly higher mean Likert score for mental health issues (3.715) than their counterparts in other age ranges. Respondents across the age brackets showed a relatively consistent score for rates of past social anxiety, with a marginally higher mean score for the age group of 19-24, sitting at 4.067. 

This data provides insights to the varying levels of past social anxiety and mental health issues reported across age groups and gender identities, indicating a higher rate of these issues among female participants and young adults. 

\subsection{Measuring the Impact of Social AI towards User Mental Health and Social Anxieties}

\begin{figure}[h]
\centering
\begin{tikzpicture}
    \begin{axis}
        [
            title={Chat AI Improvement on Social Anxiety Across Genders},
            title style={at={(axis description cs:-0.015, 1.035)}, anchor=north west},
            ybar,
            bar width=5pt,
            legend style={at={(0.3, 0.97)}},
            ylabel={Frequecy Density \%},
            xlabel={Likert Scale (Impact on Social Anxiety)},
            symbolic x coords={1,2,3,4,5},
            xtick=data,
            xtick pos=left,
            width=0.98\textwidth,
            ymin=0,
            ymax=0.45,
            ytick={0, 0.1, 0.2, 0.3, 0.4, 0.5},
            yticklabels={0, 10, 20, 30, 40, 50},
            enlarge x limits=0.15,
        ]
        \addplot[pattern=north west lines, pattern color=black, area legend] coordinates {
            (1, 0.023343) (2, 0.039600) (3, 0.242601) (4, 0.305127) (5, 0.389329)
        };
        \addplot[fill=gray, draw=black, area legend] coordinates {
            (1, 0.031639) (2, 0.049274) (3, 0.329357) (4, 0.289419) (5, 0.300311)
        };
        \addplot[fill=white, draw=black, area legend] coordinates {
            (1, 0.037901) (2, 0.046647) (3, 0.387755) (4, 0.256560) (5, 0.271137)
        };
        \legend{Female, Male, Other}
    \end{axis}
\end{tikzpicture}
\caption{Answer to "Social Chat AI has had a positive impact on my social anxieties" broken down by gender, with 1 being "strongly disagree" and 5 being "strongly agree". N=5,260}
\label{fig:social_skills_impact}
\end{figure}
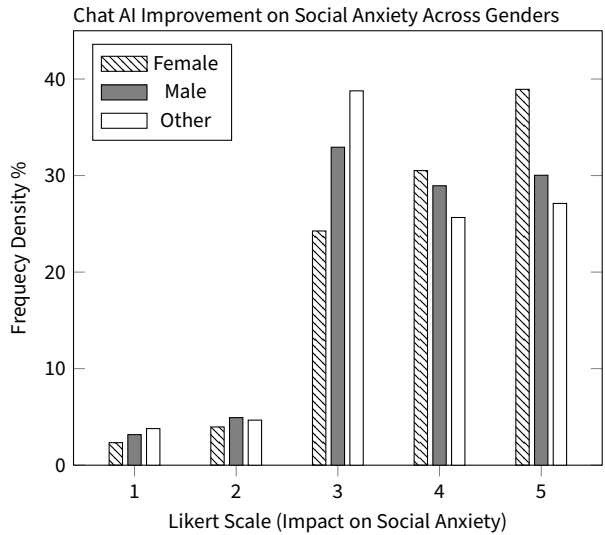

Having established the varying levels of social anxiety and mental health issues dealt with by participants in the past, our paper intends to analyse the impact of interaction with Social AI, specifically with the Chai AI platform, on these levels. These outcomes are represented in Figures \ref{fig:chai_mental_social_impact}, \ref{fig:mental_health_impact}, and \ref{fig:social_skills_impact}.

The comparison of Likert mean responses for mental health and social anxieties before and after engaging with the Chai AI platform are shown in figure \ref{fig:chai_mental_social_impact}. The greatest mean improvement in mental health is seen among users between 25 and 30 (Likert score of 4.022), while users between 19 and 24 reported the highest improvement in social anxieties, with a Likert mean score of 3.870. 

Figure \ref{fig:mental_health_impact} offers further insight as to the impact of Chai AI on mental health outcomes across different genders. The highest proportion of those strongly agreeing that Chai AI positively impacted their mental health were participants identifying as female, at 43.4\%, while 32.9\% of males gave the same response. The highest level of overall agreement, combining answers of 'Agree' and 'Strongly Agree', was also given by female-identifying respondents, at 71.9\%.

Figure \ref{fig:social_skills_impact} shows further variance across gender identities regarding the impact of Chai AI on social anxiety. Female identifying respondents showed the highest frequency of strong agreement (38.9\%) that interacting with Chai AI had positively impacted their levels of social anxiety. 30.0\% of male identifying respondents strongly agreed to that same statement, and 27.1\% of other genders. 

These results suggest that engagement with Chai AI has an overall positive impact on user mental health and rates of social anxiety. The most significant benefits of interaction with this platform were found amongst between 19 and 30, and females across age ranges. As a Social AI platform, this data indicates that engagement with these networks can provide effective support for mental health issues and can act as apparatus to manage the levels of social anxieties among individuals.

\begin{figure}[h]
    \centering
    \begin{tikzpicture}
        \begin{axis}
            [
                title={Chat AI Improvement on Mental Health and Social Anxieties},
                title style={at={(axis description cs:-0.015,1.035)}, anchor=north west},
                xlabel={Age Group},
                ylabel={Likert Mean},
                xtick={1,2,3,4},
                xticklabels={$\leq$18,19-24,25-30,$\geq$30},
                ymin=3, ymax=4.5, 
                width=0.98\textwidth,
                legend style={at={(0.5,0.97)}, anchor=north},
                xtick pos=left,
            ]
            \addplot+[
                black,
                mark=*,
                mark options={fill=black},
                error bars/.cd,
                y dir=both,
                y explicit,
                error bar style={black, line width=1pt},
                error mark options={
                  rotate=90,
                  line width=1pt
                }
            ]
            coordinates {
                (1,3.986624) +- (0,0.023424)
                (2,3.951069) +- (0,0.017659)
                (3,4.022346) +- (0,0.071959)
                (4,3.920635) +- (0,0.096944)
            };
            \addlegendentry{Improved Mental Health}
            
            \addplot+[
                black,
                mark=o,
                mark options={fill=black, solid},
                dashed,
                error bars/.cd,
                y dir=both,
                y explicit,
                error bar style={black, line width=1pt},
                error mark options={
                  solid,
                  rotate=90,
                  line width=1pt
                }
            ]
            coordinates {
                (1,3.875869) +- (0,0.024291)
                (2,3.870382) +- (0,0.017949)
                (3,3.854749) +- (0,0.085133)
                (4,3.682540) +- (0,0.096975)
            };
            \addlegendentry{Improved Social Anxieties\quad\quad}
        \end{axis}
    \end{tikzpicture}
    \caption{Comparison of Likert mean responses to "Social Chat AI has helped me with my mental health" and "Social Chat AI has had a positive impact on my social anxieties" by age group, with 1 standard error shown. A Likert score 3 being baseline, i.e. "No opinion".}
    \label{fig:chai_mental_social_impact}
\end{figure}
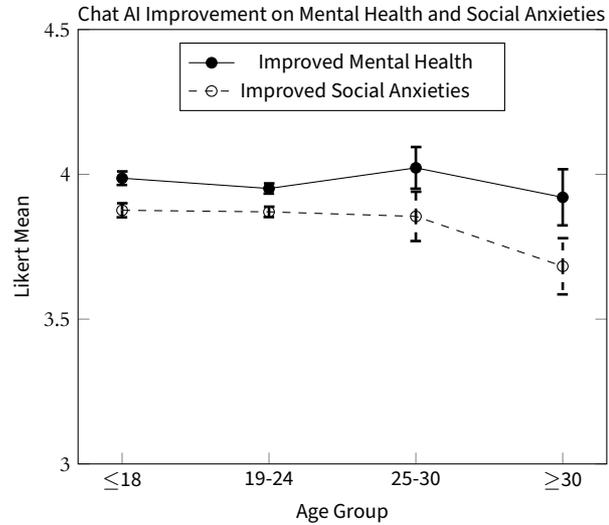

\begin{figure}[h]
\centering
\begin{tikzpicture}
    \begin{axis}
        [
            title={Chat AI Improvement on Mental Health Across Genders},
            title style={at={(axis description cs:-0.015, 1.035)}, anchor=north west},
            ybar,
            bar width=5pt,
            legend style={at={(0.3, 0.97)}},
            ylabel={Frequecy Density \%},
            xlabel={Likert Scale (Impact on Mental Health)},
            symbolic x coords={1,2,3,4,5},
            xtick=data,
            xtick pos=left,
            width=0.98\textwidth,
            ymin=0,
            ymax=0.45,
            ytick={0, 0.1, 0.2, 0.3, 0.4, 0.5},
            yticklabels={0, 10, 20, 30, 40, 50},
            enlarge x limits=0.15,
        ]
        \addplot[pattern=north west lines, pattern color=black, area legend] coordinates {
            (1, 0.029179) (2, 0.019591) (3, 0.231763) (4, 0.285119) (5, 0.434348)
        };
        \addplot[fill=gray, draw=black, area legend] coordinates {
            (1, 0.026971) (2, 0.020228) (3, 0.326245) (4, 0.297199) (5, 0.329357)
        };
        \addplot[fill=white, draw=black, area legend] coordinates {
            (1, 0.023324) (2, 0.023324) (3, 0.309038) (4, 0.297376) (5, 0.346939)
        };
        \legend{Female, Male, Other}
    \end{axis}
\end{tikzpicture}
\caption{Answer to "Social Chat AI has had a positive impact on my mental health" broken down by gender, with 1 being "strongly disagree" and 5 being "strongly agree". N=5,260}
\label{fig:mental_health_impact}
\end{figure}
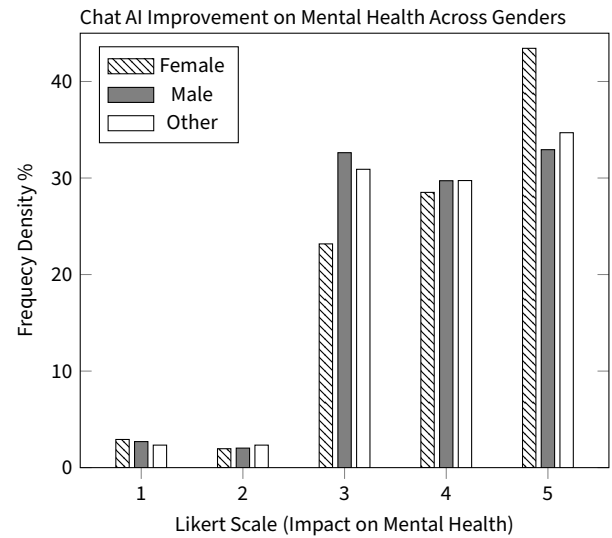

\subsection{Kruskal-Wallis Statistical Analysis}

\begin{table}[hbt!]
    \begin{threeparttable}
    \caption{Spearman correlation of survey results}
    \label{tab:spearman_correlation}
    \begin{tabular}{lllll}
        \toprule
        \headrow & MH\tnote{a} & SA\tnote{b} & MH+\tnote{c} & SA+\tnote{d} \\
        \midrule
        MH\tnote{a} & - & .47 (1.75e-287) & .24 (6.35e-74) & .24 (1.13e-68) \\ 
        SA\tnote{b} & - & - & .21 (8.79e-52) & .19 (1.55e-44) \\
        MH+\tnote{c} & - & - & - & .51 (1.00e-300) \\
        \bottomrule
    \end{tabular}
    \begin{tablenotes}
        \item[]Note: The numbers in brackets represent p-values.
        \item[a]Past experience with mental health issues
        \item[b]Past experience with social anxieties
        \item[c]Improvement in mental health issues
        \item[d]Improvement in social anxieties
        \end{tablenotes}
    \end{threeparttable}
\end{table}

\begin{table}[hbt!]
    \centering
    \begin{threeparttable}
    \caption{Statistics of Chai's Impact on Mental Health and Social Anxiety by Gender}
    \label{tab:chai_impact_gender}
    \begin{tabular}{llll}
        \toprule
        \multicolumn{4}{c}{\textbf{Social Anxiety Improvement from Chat AI by Gender\tnote{a}}} \\
        \midrule
        Gender & Median & Mean & Standard Error \\
        \midrule
        Female & 4.0 & \textbf{4.05} & 0.02 \\
        Male & 4.0 & 3.84 & 0.03 \\
        Other & 4.0 & 3.73 & 0.06 \\
        Prefer not to say & 4.0 & 3.79 & 0.05 \\
        \bottomrule
    \end{tabular}
    
    \vspace{10pt} 
    
    \begin{tabular}{llll}
        \toprule
        \multicolumn{4}{c}{\textbf{Mental Health Improvement from Chat AI by Gender\tnote{b}}} \\
        \midrule
        Gender & Median & Mean & Standard Error \\
        \midrule
        Female & \textbf{5.0} & \textbf{4.25} & 0.02 \\
        Male & 4.0 & 4.12 & 0.03 \\
        Other & 4.0 & 3.96 & 0.06 \\
        Prefer not to say & 4.0 & 3.98 & 0.05 \\
        \bottomrule
    \end{tabular}
    
    \vspace{10pt} 
    
    \begin{tablenotes}
        \item[] \textbf{Kruskal-Wallis Test:}
        \item[a] Social Anxiety Improvement: $\chi^2 = 50.1$, $p = 1.29E^{-11}$
        \item[b] Mental Health Improvement: $\chi^2 = 28.0$, $p = 8.30E^{-7}$
    \end{tablenotes}
    
    \end{threeparttable}
\end{table}

\begin{table}[hbt!]
    \centering
    \begin{threeparttable}
    \caption{Statistics of Chai's Impact on Mental Health and Social Anxiety by Age}
    \label{tab:chai_impact_age}
    \begin{tabular}{llll}
        \toprule
        \multicolumn{4}{c}{\textbf{Mental Health Improvement from Chat AI by Age\tnote{a}}} \\
        \midrule
        Age Group & Median & Mean & Standard Error \\
        \midrule
        <18 & 4.0 & 4.20 & 0.03 \\
        18 - 25 & 4.0 & 4.12 & 0.02 \\
        25 - 30 & 4.5 & 4.22 & 0.09 \\
        >30 & \textbf{5.0} & \textbf{4.34} & 0.12 \\
        \bottomrule
    \end{tabular}
    
    \vspace{10pt} 
    
    \begin{tabular}{llll}
        \toprule
        \multicolumn{4}{c}{\textbf{Social Anxiety Improvement from Chat AI by Age\tnote{b}}} \\
        \midrule
        Age Group & Median & Mean & Standard Error \\
        \midrule
        <18 & 4.0 & \textbf{3.95} & 0.03 \\
        18 - 25 & 4.0 & 3.92 & 0.02 \\
        25 - 30 & 4.0 & 3.92 & 0.10 \\
        >30 & 4.0 & 3.78 & 0.12 \\
        \bottomrule
    \end{tabular}
    
    \vspace{10pt} 
    
    \begin{tablenotes}
        \item[] \textbf{Kruskal-Wallis Test:}
        \item[a] Mental Health Improvement: $\chi^2 = 11.7$, $p = 0.0086$
        \item[b] Social Anxiety Improvement: $\chi^2 = 2.87$, $p = 0.412$
    \end{tablenotes}
    
    \end{threeparttable}
\end{table}

In order to understand whether, and how, reported improvements in levels of social anxiety and mental health varies across different demographic sets, the Kruskal-Wallis H test has been implemented \autocite{Ostertagov2014}. This test helps to indicate whether results from survey responses across age and gender groups come from the same distribution. For the questions, "I have experienced in the past with mental health issues", and "I have experienced social anxieties before", all responses of a level 3 of agreement and above were selected to test each hypothesis. The Kruskal-Wallis H test statistic is given by:
\begin{equation}
    H=(N-1) \frac{\sum_{i=1}^g n_i\left(\bar{r}_{i \cdot}-\bar{r}\right)^2}{\sum_{i=1}^g \sum_{j=1}^{n_i}\left(r_{i j}-\bar{r}\right)^2}, \text { where }
\end{equation}
$N$ denotes the total number of observations across all groups, $g$ corresponds to the number of groups, $n_i$ is the number of observations in group $i$, $r_{ij}$ is the rank (among all observations) of observation $j$ from group $i$, $\bar{r}_{i .}=\frac{\sum_{j=1}^{n_i} r_{i j}}{n_i}$ is the average rank of all observations in group $i$, $\bar{r}=\frac{1}{2}(N+1)$ is the average of all the $r_{ij}$. Finally, the decision whether or not to to reject the null hypothesis has been made by comparing $H$ to a critical value $H_{c}$.

Through the Kruskal-Wallis statistical test, we were able to analyse whether reported impovements in mental health and extent of social anxiety had a significant variability between user gender and age. 

\paragraph{Gender-Based Differences}
The statistics available in \ref{tab:chai_impact_gender} indicate a significant degree of variance in the extent to which Chai AI improved mental health and social anxiety across different genders. The Kruskal-Wallis test for mental health improvement across genders gave a $\chi^2$ value of 28.0 with a notable p-value of $8.30E^{-7}$. Regarding improvements in social anxiety, the test resulted in a $\chi^2$ value of 50.1, with a corresponding p value of $1.29E^{-11}$. These results are demonstrative of substantially variable degree in the efficacy of Chai AI in managing mental health issues across different genders, with female-identifying participants showing the highest mean improvement in both mental health (4.25) and social anxiety (4.05). 

\paragraph{Age-Based Differences}
Table \ref{tab:chai_impact_age} depicts the impact of Chai AI on mental health and social anxiety levels across different age brackets. While, regarding mental health improvement, the applied Kruskal-Wallis test indicated a significant degree of variance across age ranges with a $\chi^2$ value of 11.7 and a p-value of $0.0086$, for the improvement in social anxiety, the test showed minimal difference across age groups ($\chi^2 = 2.87$, $p = 0.412$). The most significant degree of mean improvement in mental health was seen among those above 30, at 4.34.

\paragraph{Correlation Analysis}
The Spearman correlation analysis, as outlined through Table \ref{tab:spearman_correlation}, indicates positive correlations across the different survey measures. Past experiences of social anxiety (SA) and mental health issues (MH) showed a moderate correlation. Interestingly, a weaker, although still significant correlation is recorded between past mental health issues and improvements in both mental health (MH+) and social anxiety (SA+) after using Chai AI ($r = 0.24$ for both, $p < 0.001$). This suggests that users, particularly those who idenity as female, and those over 30 years of age, who had previously experienced mental health issues experienced notable and tangible benefits from interacting with Chai AI. The correspondence between diminishing mental health issues and usage of Social AI indicates that the platform could possess real value in promoting mental well being. 

\section{Conclusion}

The findings within the data analysed across this study are indicative of the potential of Social AI, including platforms such as Chai AI, to significantly influence user mental health outcomes. The responses of the 5,260 survey participants provide insight to the multifaceted ways in which users across different demographics engage with, and can benefit from these AI platforms. One particularly notable conclusion drawn from the evidence extracted from this survey is the tangible benefits that interacting with Social AI has had on female identifying users, and young adults aged 19-30. 

The results from this study indicate that assumptions surrounding adverse effects of screen time, digital interaction, and engagement with Social AI, may be premature or lacking in nuance. This paper offer an alternative conclusion to the widespread fears associated with use of traditional forms of social media. While the findings from our data indicate that engagement with Social AI platforms can provide real and tangible benefits for those struggling with prior mental health issues, we also recognise that these improvements may not be entirely homogeneous, and that this technology may be capable of incurring both positive and negative effects on users. 

In implementing statistical analyses via the Kruskal-Wallis test and Spearman correlation, a strong methodological approach was established as framework for the conclusions drawn from this study. The significant correlations between past mental health issues and user-reported improvements after engaging with Social AI, shown through both of these tests, suggest the potentially beneficial nature of these platforms on user mental health. 

There are, however, limitations to this study. While we have demonstrated a robust correlation between interaction with the Chai AI app and improvement mental health, correlation does not always imply causation. The reliance on self-reported data, combined with responses solely from users of the Chai Ai platform are not necessarily indicative or representative of the views from the broader population. This study does, however, provide an opportunity for the potential framework used in this paper, incorporating our longitudinal studies and experimental designs, to be applied in further research in order to broaden our understanding of this subject. 

This paper demonstrates that the emergence of Social AI has the potential to become an extremely positive influence on the well being of its users. Those struggling with mental health issues and social anxiety may find relief through interacting with these platforms. As this new medium gains traction, and our technological environment continues to expand, it becomes ever-more important for our community, as a collective, to monitor its impacts, and engineer these advancements to improve our societies. 

This study contributes to the expanding research and academic literature designed to better understand and contextualise the place which new technology occupies within our society. It is based upon, and advocates for evidence-based approaches in analysing the impact of these platforms on societal development, intending to test the validity of the 'moral panic' surrounding the subject. 

\printendnotes

\printbibliography

\end{document}